\documentstyle[12pt,aaspp4]{article}
\def\etal{{et al.}}
\def\cf{{c.f.}}

\def\deg{{^\circ}}

\slugcomment{POPe-732, KSUPT-97/4, KUNS-1469 \hspace{0.3truecm} November 1997}

\lefthead{Park et al.}
\righthead{CMB Anisotropy Maps}  

\begin{document}

\title{CMB Anisotropy Correlation Function and Topology from Simulated Maps 
for MAP}

\author{Changbom Park\altaffilmark{1}, 
        Wesley N. Colley\altaffilmark{2,5}, 
        J. Richard Gott III\altaffilmark{2},
        Bharat Ratra\altaffilmark{3}, 
        David N. Spergel\altaffilmark{2},
        and
        Naoshi Sugiyama\altaffilmark{4}}

\altaffiltext{1}{Department of Astronomy, 
   Seoul National University, Seoul 151. cbp@astro.snu.ac.kr}
\altaffiltext{2}{Department of Astrophysical Sciences, 
   Princeton University, Princeton, NJ 08544.}
\altaffiltext{3}{Department of Physics, Kansas State University, 
   Manhattan, KS 66506.}
\altaffiltext{4}{Department of Physics, Kyoto University,
  Kitashirakawa-Oiwakecho, Sakyo-ku, Kyoto 606.}
\altaffiltext{5}{Current Address:  Harvard-Smithsonian Center for Astrophysics,
   Cambridge, MA 02138.}

\begin{abstract}
We have simulated cosmic microwave background (CMB) anisotropy maps for several
$COBE$-DMR-normalized cold dark matter (CDM) cosmogonies, to make predictions 
for the MAP experiment, an upcoming whole-sky CMB anisotropy space mission. We 
have studied the sensitivity of the simulated MAP data to cosmology, sky 
coverage, and instrumental noise. Given an accurate knowledge of instrumental 
noise,  MAP data will ably discriminate amongst the cosmogonies considered, and 
superbly determine the topology of the initial fluctuations. 

A correlation function analysis of the simulated MAP data results in a very
accurate measurement of the acoustic Hubble radius at decoupling. A low-density 
open CDM
model with $\Omega_0=0.4$ can be distinguished from the $\Omega_0=1$ fiducial
CDM model or a spatially-flat CDM model with a cosmological constant and
$\Omega_0=0.4$ with more than 99\% confidence from the location of the acoustic
``valley" in the correlation function. 

A genus analysis of the simulated MAP data indicates that in cosmogonies
with Gaussian random-phase initial conditions,
the horizontal shift of the zero-crossing point of the genus curve 
near the mean temperature threshold level $\nu=0$ should not exceed 
$|\Delta\nu| = 0.01$ (0.04) when the total effective FWHM smoothing is 
$0.3\deg$ $(1.0\deg)$. The asymmetry of the genus curve at the positive and 
negative threshold levels should not exceed $|\Delta g/g(\nu=\pm 1)| = 
0.8\%$  (4\%) at $0.3\deg$ $(1\deg)$ FWHM smoothing. Deviations of the observed 
MAP data in excess of these small values will be evidence for non-Gaussian 
behavior.

The amplitude of the genus curve is a measure of the shape of the power
spectrum at the smoothing scale.  Even with the expected amount of instrumental 
noise and partial sky coverage (due to the Galaxy), the MAP data should 
allow discrimination amongst the cosmogonies considered at more than 99\% 
confidence solely from a genus amplitude analysis.
\end{abstract}


\section{Introduction}

Following the $COBE$-DMR detection of anisotropy in the cosmic microwave
background on large angular scales (Smoot et al. 1992; Wright et al. 1992; 
Bennett et al. 1996; G\'orski et al. 1996), there have been many measurements 
of CMB anisotropy on angular scales down to $\sim 10'$ (Ganga et al. 1994; 
Guti\'errez et al. 1997; Piccirillo et al. 1997; Netterfield et al. 1997; 
Gundersen et al. 1995; Tucker et al. 1997; Platt et al. 1997; Masi et al. 1996; 
Lim et al. 1996; Cheng et al. 1997; Griffin et al. 1997; Scott et al. 1996; 
Leitch et al. 1997; Church et al. 1997, see Page 1997 for a review). In 
addition to the DMR experiment, CMB anisotropy maps
have been produced from four other experiments (Ganga et al. 1994; White \&
Bunn 1995; Scott et al. 1996; Tegmark et al. 1997). CMB anisotropy observations
are beginning to test cosmogonical models and provide interesting constraints
on cosmological parameters in these models (e.g., Bunn \& Sugiyama 1995;
G\'orski et al. 1995, 1998; Stompor, G\'orski, \& Banday 1995; Ganga, Ratra, \&
Sugiyama 1996; Ganga et al. 1997a,b, 1998; Hancock et al. 1997; Bond \& Jaffe
1997; Stompor 1997; Lineweaver \& Barbosa 1997; Ratra et al.  1997b). However, 
definitive results, including significant constraints on
cosmological parameters, must await new CMB anisotropy data, covering a large
fraction of the sky (to minimize sample variance) and acquired at a variety of
frequencies (to allow for control over possible non-CMB anisotropy foreground
contamination).  Of the currently available data sets, the DMR maps best meet
these criteria.

Plans are afoot to launch two second-generation CMB anisotropy satellite
missions --- the Microwave Anisotropy Probe (MAP)\footnote{ 
The MAP homepage is at http://map.gsfc.nasa.gov/} 
is expected to have data in less than four years, and the Planck 
Surveyor\footnote{ 
The Planck Surveyor homepage is at
http://astro.estec.esa.nl/SA-general/Projects/Cobras/cobras.html} 
is expected to have data in less than a decade. These missions are expected to 
measure the CMB anisotropy on angular scales larger than a fraction of a degree, at a variety of frequencies.  MAP will probe the anisotropy at frequencies of 
22, 30, 40, 60, and 90 GHz at expected angular resolutions of 0.93, 0.68, 0.47,
0.35, and 0.21 degrees (FWHM beamwidths), respectively. At 90 GHz the MAP 
sensitivity is expected to be $\sim 35\ \mu$K per $0.3^\circ\times 0.3^\circ$ 
pixel. Planck Surveyor is expected to measure the CMB anisotropy at frequencies 
ranging from 30 to 900 GHz with angular resolution ranging from 0.5 to 0.075 
degrees.

It is now possible to make accurate predictions for the CMB anisotropy, as a
function of cosmological parameters, in many different cosmological models.  In
conjunction with such theoretically computed model CMB anisotropy spectra, the
data from these experiments should lead to an accurate determination of
cosmological parameters, such as the mass density parameter $\Omega_0$, the
baryonic-mass density parameter $\Omega_B$, the cosmological constant ${\Lambda}$, the Hubble parameter $H_0$, the matter power spectrum, and the
optical depth to the last scattering surface (e.g., Spergel 1994; Jungman et al. 1996; Bond,
Efstathiou, \& Tegmark 1997; Zaldarriaga, Spergel, \& Seljak 1997; Dodelson,
Kinney, \& Kolb 1997). At the least, the data from these space missions should
in principle be able to rule out most cosmological models currently under
consideration. The present situation in cosmology --- where one is able to make
definite theoretical predictions which can be judged by near-future
observations --- is quite unusual.

Prior to the $COBE$-DMR measurement of the large-scale CMB anisotropy,
Gott et al. (1990) had generated simulated all-sky CMB anisotropy maps
predicted in the spatially-flat $\Omega_0 = 1$ fiducial CDM model with Gaussian, adiabatic, scale-invariant matter fluctuations. The DMR data is indeed consistent with such a CMB anisotropy spectrum (G\'orski et al. 1996, 1998).

In this paper we generate simulated all-sky CMB anisotropy maps appropriate for
MAP in several representative cosmological models normalized to the DMR data.
Our predictions can be directly compared to the MAP data when it is available.  
Hinshaw, Bennett, \& Kogut (1995) have generated all-sky
CMB anisotropy maps expected in an $\Omega_0=1$ CDM model, at angular
resolutions of $0.5^\circ$ and $1^\circ$, to study the effects of sky coverage,
angular resolution, instrumental noise, and unresolved features in the data.
Novikov \& J{\o}rgensen (1996) have simulated $10^\circ\times 10^\circ$ sky
anisotropy maps of $\Omega_0=1$ CDM models with $\Omega_B=0.1$ and 0.03
at $0.5^\circ$--$1^\circ$ 
resolutions.  They have examined various statistical properties of the
temperature fluctuation field, such as clustering of peaks and percolation.  We
focus here on generating mock CMB anisotropy maps appropriate for the specific
experimental configuration of MAP.  We also use these CMB anisotropy maps
to examine the ability of the expected MAP data to discriminate between
different cosmological models. In what follows we assume that the MAP data will
be limited by statistical noise rather than foregrounds or systematic errors.

In Section 2 we present the cosmological models we study, and explain how the
simulated CMB anisotropy maps are generated.  In Section 3 we compute the
CMB anisotropy correlation function for each of the simulated maps, and show 
that it is sensitive
to the acoustic peaks in the CMB anisotropy angular power spectrum.  In Section
4 we study the genus statistic of the generated maps, and explore its ability to
discriminate between cosmological models and to detect non-Gaussian
fluctuations in the CMB anisotropy field.  The effect of instrumental noise is
considered.  Conclusions follow in Section 5.

\section{Simulation of CMB Anisotropy Maps}

To generate simulated CMB anisotropy maps for MAP we have chosen four
representative Gaussian, adiabatic cosmogonical models, with distinct CMB
anisotropy angular power spectra: (1) The fiducial CDM model (FCDM) with
$\Omega_0=1$, the Hubble parameter $H_0$ divided by 100 km/sec/Mpc $h = 0.5$, 
and with no reionization. (2) A spatially-flat, low-density CDM model 
($\Lambda$CDM) with a cosmological constant $\Lambda$, $\Omega_0=0.4$, and 
$h=0.6$. (3) A spatially-open low-density CDM model (OCDM) with no $\Lambda$, 
$\Omega_0=0.4$, and $h=0.65$. (4) A reionized fiducial CDM model (RCDM) with
$\Omega_0 = 1$, $h = 0.5$, and fully reionized at redshift $z = 100$ (i.e., 
with electron ionization fraction $x_e = 1$ at $z \leq 100$ and so
optical depth $\tau =1$). For all models, the
baryonic-mass density parameter is $\Omega_B =0.0125 h^{-2}$. Further 
details are given in Ratra et al. (1997a) and Sugiyama (1995).

The FCDM and $\Lambda$CDM models considered here assume a scale-invariant
primordial energy-density power spectrum (Harrison 1970; Peebles \& Yu 1970;
Zel'dovich 1972), as is predicted in the simplest spatially-flat inflation
models (Guth 1981, also see Kazanas 1980; Sato 1981).  The FCDM model is
inconsistent with galaxy clustering data (e.g., Maddox et al. 1990; Saunders et 
al. 1991; Vogeley et al. 1992; Park et al. 1994), but still serves as a useful
fiducial for cosmological model comparison.  $\Lambda$CDM models (Peebles
1984) became popular when it was discovered that the large-scale galaxy 
distribution was of the form expected in CDM models with $\Omega_0 h \sim 0.2 - 
0.3$ (e.g., Park 1990, 1991; Vogeley et al. 1992; da Costa et al. 1994; 
Peacock \& Dodds 1994), since
these models could be accommodated in the then-dominant spatially-flat inflation
scenario.  However, an open inflating universe can also be created as a single 
bubble during the decay of an initial metastable inflation state (Gott 1982, 
1986). The OCDM model considered here assumes the
primordial energy-density power spectrum of the simplest open-bubble inflation 
models (Ratra \& Peebles 1994, 1995; Bucher, Goldhaber, \& Turok 1995; 
Yamamoto, Sasaki, \& Tanaka 1995). The low-density models, OCDM and 
$\Lambda$CDM, are consistent with most current large-scale structure and CMB 
anisotropy observations (e.g., Gott 1997; Turner 1997; Cole et al. 1997; 
Gardner et al. 1997; Croft et al. 1997; Coles et al. 1997; Jenkins et al. 1997).

A CMB temperature fluctuation field on the surface of a sphere (i.e., the sky)
can be decomposed into a sum of spherical harmonics, ${\delta T/ T}(\theta, 
\phi) = \sum a^m_{\ell} Y_{\ell}^{m}(\theta, \phi)$, where $(\theta, \phi)$ are
the angular coordinates on the sky. If the initial matter fluctuation is a 
Gaussian random field, as predicted in simple inflation models, the expansion
coefficients $a^m_{\ell}$ are independent Gaussian random variables with zero
means and variances $\langle |a^m_{\ell}|^2 \rangle = C_{\ell}$, where
$C_{\ell}$ is the CMB anisotropy angular power spectrum (see below for more 
details).

The CMB anisotropy angular power spectra of the four models considered 
here are shown in Figure 1 (Ratra et al. 1997a; Sugiyama 1995). Following 
G\'orski et al. (1998) and Stompor (1997), the angular power spectra are
normalized to the arithmetic mean of the $\pm$2-$\sigma$ $COBE$-DMR 
53 and 90 GHz data from the two extreme data sets: (1) galactic-frame
maps accounting for the high-latitude Galactic emission correction and 
including the $\ell = 2$ moment in the analysis; and (2) ecliptic-frame
maps ignoring the high-latitude Galactic emission correction and excluding
the $\ell = 2$ moment from the analysis. 

At large angular scales the CMB anisotropy angular power spectra in the four
models are of approximately the same amplitude. At smaller angular scales
acoustic oscillations due to the pressure of the photon-baryon fluid and matter
velocity perturbations at photon decoupling cause the angular power spectra to
rise. At still smaller angular scales photon diffusion and the finite thickness
of the last-scattering surface smears out the fluctuations. These effects give
rise to the characteristic acoustic peaks in the angular power spectra on
intermediate angular scales. In an open model the geometrical effect that the
circumference of a circle is larger than $2\pi$ times its radius moves these
peaks to smaller angular scales relative to the spatially-flat case
(Kamionkowski, Spergel, \& Sugiyama 1994). In the
spatially-flat $\Omega_0 = 1$ FCDM and low-density $\Lambda$CDM models, the
first acoustic peak is at $\ell = 217$ and $\ell = 222$, respectively. The
amplitudes of the peaks in these models are different.  In the open case the
first peak is at $\ell = 349$, while in the reionized RCDM model the peaks are
smeared out. Consequently, the presence, location, and amplitude of the
acoustic peaks are a valuable probe of the geometry and content of the
universe.

In real observations and analyses, the temperature fluctuations on the sky are
convolved with the beam pattern of the experiment, and are often further
smoothed to increase the signal-to-noise ratio.  If the beam pattern is
$B(\theta)$ and the smoothing filter is $F(\theta)$, the observed CMB
anisotropy temperature field is
\begin{equation}
{\delta T\over T}(\theta,\phi) =
   \sum_{\ell=2}^{\infty}\sum_{m=-\ell}^{\ell} a^m_{\ell}
   Y_{\ell}^{m}(\theta,\phi)B_{\ell} F_{\ell},
\end{equation}
where $B_{\ell}$ and $F_{\ell}$ are the coefficients of the beam pattern and
the smoothing filter expanded in Legendre polynomials, and we consider
only harmonics higher than the dipole. Assuming the idealized case of an
all-sky data map, the inverse relation for the harmonic amplitude is
\begin{equation}
a_{\ell}^m = \int d\Omega {\delta T\over T}(\theta,\phi)
     Y_{\ell}^m(\theta,\phi)^* /B_{\ell}F_{\ell}.
\end{equation}
For a Gaussian beam (or a smoothing filter) with a width 
$\lambda\ll 1$ to a very good approximation,
\begin{equation}
B_{\ell} = \exp(-(\ell+1/2)^2 \lambda^2/2).
\end{equation}

We use equation (1) to simulate the CMB anisotropy maps for MAP.
We limit the summation to $\ell \leq 1500$. For each cosmogony the
real and imaginary components of $a^m_{\ell}$ are randomly sampled Gaussian
variates with zero means, and variances equal to $C_{\ell}/2$ for the $m \neq
0$ harmonics and equal to $C_{\ell}$ for the $m=0$ harmonics.  We adopt a
Gaussian beam with FWHM $\theta_B=0.21^\circ$ (or width $\lambda=\theta_B
/2\sqrt{2\ln 2}$ in equation [3]).  The temperature fluctuations are then
computed at $1801\times 3600$ points in ($\theta, \phi$) space on a grid with
$0.1^\circ$ spacing.

The data from MAP is expected to include instrumental noise of $\sim 35$ 
$\mu$K per $0.3^\circ\times 0.3^\circ$ pixel.  If the noise is white, its power
spectrum is constant $C_{\ell}\equiv C_N$.  Consider a pure noise map convolved
with a top hat filter $F$ with area of 0.09 square degrees. Then the variance
of the noise temperature fluctuation, which is set to $(35\ \mu$K$/T_0)^2$ per
0.09 square degrees, is given by equation (5) below (at $\theta=0$
with $C_{\ell}$ replaced by
$C_N$). The top hat filter is $F(\theta)=1/\pi\theta_H^2$ for
$\theta\leq\theta_H$, and $0$ otherwise.  The coefficients of its expansion
into Legendre polynomials are, for $\theta_H\ll 1$,
\begin{equation}
  F_{\ell}=\sum_{n=0}^{\infty}{(-)^n\over (n!)^2(n+1)}
  [(\ell+1/2)\theta_H/2]^{2n}.
\end{equation}
We obtain from equation (5) $C_N=5\times 10^{-15}$ for MAP.  One can study the
effects of noise on statistics using this noise power spectrum.  In practice,
of course, the MAP results will include an estimate of the noise at each pixel
derived from the many observations taken as the satellite's horns sweep
out complicated paths on the sky.  The satellite horns are paired, so 
subtracting their signals gives the noise map; this allows an extremely 
accurate characterization of the noise as a function of position on the sky.
Such a detailed noise map will allow a much more careful accounting of 
instrumental noise. (For example, the noise at the ecliptic poles is expected to be about half of that at the ecliptic plane).

The maps resulting from the $\Lambda$CDM and OCDM models with simulated 
(constant over the sky) instrumental noise, projected according to 
the Mollweide projection method (Snyder 1993), are shown in Figure 2 (Plate 1).  These maps are smoothed with a $0.214\deg$ FWHM Gaussian, which makes the total 
smoothing scale of the signal $0.3\deg$.
The Mollweide equal area projection shows the whole sky in a single map,
but structures are distorted.  In the topology study we use conformal
stereographic projection maps where shapes of structures are preserved locally.
Stereographic maps of the CMB fluctuations simulated for the OCDM and 
$\Lambda$CDM models and convolved with a $0.978\deg$ FWHM Gaussian (which 
makes the total smoothing scale of the signal $1.0\deg$) are shown in Figure 3 
(Plate 2). At this smoothing length most structures in the map 
generated by MAP will correspond to actual CMB anisotropy fluctuations (see 
Figure 8). It is to be noted that the OCDM and $\Lambda$CDM maps are visually
very similar --- this is a consequence of the expected instrumental noise and
smoothing (see Figures 1$b$ and 1$c$). We emphasize, however, that this does 
not mean that MAP can not statistically distinguish between these models.

\section{Correlation Function}

Differences between cosmological models are often displayed in terms of the
angular power spectrum, as in Figure 1. Most models have relatively small
differences at large scales ($\ell \leq 30$), but they often have relatively
large differences at small scales ($\ell \geq 100$).  The structure of the
acoustic peaks at smaller angular scales depends on the cosmogony and the
cosmological parameters.  Hence, to test cosmogonies and measure these
parameters, it is essential to have an accurate measurement of the location and
amplitude of the acoustic peaks. Note that the acoustic peaks cause an
oscillatory wave in the power spectrum (see Figure 1). Furthermore, the
amplitudes of individual harmonics in the power spectrum are destined to 
fluctuate with respect to the mean. To measure precisely 
the amplitude and location of the acoustic peaks in the power spectrum one
must make a fit over many harmonics.

An important integral of the power spectrum is the two-point
auto-correlation function, the real-space Fourier counterpart of the power
spectrum.  Since the hot and cold spots due to the acoustic peaks in the power
spectrum are compact in real space (their size is fixed by the acoustic Hubble 
radius at decoupling --- see Figures 2 and 3), it is useful to explore them in
real space.  (Note that in Fourier space the peaks are well-approximated by a 
wavy sinc-function.) While the correlation function and the power spectrum are
simply related, it is useful to analyze data both in real space and in Fourier 
space (e.g., Hinshaw et al. 1996; G\'orski et al. 1996; Wright et al. 1996).

The correlation function of CMB temperature fluctuations, 
corresponding to equation (1), is
\begin{equation}
C(\theta)={1\over 4\pi}\sum_{\ell=2}^{\infty}
  C_{\ell} (2\ell+1) P_{\ell}(\cos \theta) B_{\ell}^2 F_{\ell}^2 .
\end{equation}
Here $P_{\ell}$ are Legendre polynomials and we have ignored the monopole
and dipole harmonics. The power spectra shown in Figure 1$a$ can be 
approximated by an exponential minus a sinc-function. For example, 
the acoustic peaks of the FCDM power spectrum can be approximately
described by the function $-{\rm sinc}(4.5\ell/\ell_{max})$. 
Consequently, these acoustic peaks should
appear as a negative top hat in the correlation function.

Figure 4 shows the correlation functions of the four models considered.  Thick
curves are the correlation functions transformed from the power spectrum curves
in Figure 1.  Thin solid curves, tracing the thick ones at large angular 
separations, account for the convolution with a $0.21^\circ$ Gaussian beam in
Figure 4$a$ and with additional $0.214^\circ$ Gaussian smoothing in Figure
4$b$. The acoustic peaks in the power
spectrum indeed appear as a valley in this plot. The position, depth, and width
of the acoustic valley depend on the model and the cosmological parameters. We
have generated seven sets of $a_{\ell}^m$'s for the FCDM model, and have
computed their correlation functions. These are also plotted, using dots, in
Figure 4. At separations $\geq 1^\circ$ cosmic variance causes a large scatter 
amongst the seven correlation functions. Hence it is very difficult to
discriminate between models using large angular scale data alone. However,
direct measurement from these realizations shows that the FCDM acoustic valley 
in the function $\theta^{1/2}C(\theta)$ is consistently located at 
$\theta=1.13^\circ\pm 0.02^\circ$, comparing excellently with the theoretically
predicted FCDM acoustic valley location at $\theta=1.13^\circ$ for a 
$0.21^\circ$ beam (and at $1.17^\circ$ in the unsmoothed map).
The OCDM model acoustic valley lies at a smaller separation of $0.65^\circ\pm 
0.03^\circ$ ($0.69^\circ$ in the unsmoothed map) --- clearly distinguishable from the acoustic valleys of the FCDM and $\Lambda$CDM models (some 13-$\sigma$ 
difference for the fixed cosmological parameter values used here).

On the other hand, the spatially-flat FCDM and $\Lambda$CDM models have 
acoustic peaks at nearly the same angular scales. Therefore, they have to be differentiated by
the amplitude of the power spectrum or the correlation function.  It can be
noted from Figure 4 that the correlation functions of the FCDM and $\Lambda$CDM
models are statistically significantly different at separations $\theta\approx
0.5^\circ$. In the case of an ideal sample with whole-sky coverage and with no
noise, the MAP data would distinguish the correlation function of the FCDM
universe from that of the $\Lambda$CDM case at the 2.6-$\sigma$ level at
angular separations near $\theta=0.5^\circ$. Note that this ignores the
relative uncertainty between the DMR normalizations of the FCDM and 
$\Lambda$CDM models. The numerical value of this relative uncertainty is
not known, however, the absolute model-dependent uncertainty in the DMR 
normalization is $\sim 10$--$12\%$ (in $Q_{\rm rms-PS}$, the quadrupole amplitude 
of the CMB anisotropy, G\'orski et al. 1998). Also note that, for this 
comparison between FCDM and $\Lambda$CDM, cosmological parameters such as 
$h$ and $\Omega_B$ are held fixed. 

To estimate the effects of instrumental noise on the correlation 
function analysis of the MAP data we use the noise power 
spectrum, $C_N = 5\times 10^{-15}$, derived above.
This noise power is added to the CMB anisotropy signal, and
the map is then further smoothed with a $0.214\deg$ FWHM Gaussian.
That is, we generate a set of $(a_{\ell}^m B_{\ell}+n)F_{\ell}$ up to 
$\ell=2500$ where $n$ is a Gaussian random variable with zero mean and
$\langle n^2\rangle = C_N$, and $B_{\ell}$ and $F_{\ell}$
represent the $0.21\deg$ beam and $0.214\deg$ smoothing filter,
respectively.  The corresponding correlation function is then computed. 
From seven realizations of the noise-added correlation
functions for FCDM and OCDM the positions of the acoustic valleys are 
measured to be at $\theta=1.13\deg\pm 0.03\deg$ and $0.65\deg\pm 0.04\deg$,
respectively.  The correlation function acoustic valleys of the FCDM and OCDM 
models are now 9-$\sigma$ apart. It is also noted (from Figure 4) that 
the statistical uncertainty in the amplitude of correlation function at 
separations $\theta\ge 0.5\deg$ is hardly affected by the instrumental noise 
expected for MAP.  This means that cosmic variance, rather than instrumental 
noise, dominates the statistical uncertainty in the correlation function at 
these scales. (This is also true for the genus statistic --- see Table 1.)  Therefore, the above argument on discrimination between the spatially-flat 
models on the basis of the amplitude of the correlation function remains 
essentially unchanged.

\section{Topology}

The power spectrum and correlation function measured from the observed CMB
anisotropy maps provides important information on the nature of density
fluctuations in the early universe, thus constraining structure formation
mechanisms.  Another important statistical characteristic of a CMB anisotropy
map is whether or not its temperature fluctuation field is Gaussian 
random-phase.  One can test for Gaussianity of the temperature distribution by 
constructing histograms of the pixel values. To test the Gaussianity 
and phase properties of the maps the genus topology test is useful.  In 
this section we study the topology of the CMB anisotropy isotemperature contour 
surfaces.  This provides information on the global structure of the density 
fluctuation field.

\subsection{The Genus}

The genus has been used as a measure of topology of the matter/galaxy 
distribution. For three-dimensional topology see Gott, Melott, \& Dickinson 
(1986), Gott, Weinberg, \& Melott (1987), Vogeley et al. (1994), and Cole 
et al. (1997); for two-dimensional topology see Melott et al. (1989), 
Park et al. (1992), and Colley (1997). The genus has also been used as a 
measure of the topology of CMB anisotropy fields (Coles 1988; Park \& Gott
1988; Gott et al. 1990; Colley, Gott, \& Park 1996; Kogut et al. 1996;
Schmalzing \& G\'orski 1997; Winitzki \& Kosowsky 1997). For the
two-dimensional CMB anisotropy temperature field, the genus is the number of
hot spots minus the number of cold spots, i.e., at a temperature threshold
level $\nu$ the genus is
\begin{equation}
     g(\nu) = {1\over 2\pi} \int_C \kappa ds,
\end{equation}
where $\kappa$ is the signed curvature of the isotemperature contours
$C$.  The genus curve as a function of the temperature threshold level
has a characteristic shape for a Gaussian random-phase field. Changing the shape
of the CMB anisotropy power spectrum only changes the amplitude of the 
genus curve. Hence the genus curve provides a good measure of both 
non-random phases and the shape of the power spectrum. The above real-space 
definition of the genus allows it to be directly used for studying the 
statistics of the acoustic peak hot and cold spots. Previous measurements of the
CMB anisotropy genus (see references above) were consistent with the Gaussian random-phase hypothesis.

The theoretical  genus per steradian of a two dimensional Gaussian
field with correlation function $C(\theta)$ is (Gott et al. 1990)
\begin{equation}
g(\nu) = {1\over (2\pi)^{3/2}} {C^{(2)}\over C^{(0)}}\nu e^{-\nu^2/2},
\end{equation}
where $C^{(n)}\equiv (-)^{n/2}(d^n C/d\theta^n)_{\theta=0}$, and the
threshold temperature fluctuation is $\nu \sqrt{C^{(0)}} = \nu\sigma$.
For a CMB anisotropy temperature map convolved with beam $B$ and smoothing
filter $F$, $C^{(0)}$ and $C^{(2)}$ are given by
\begin{eqnarray}
C^{(0)} &=& {1\over 4\pi}\sum (2\ell+1) C_{\ell} B_{\ell}^2 F_{\ell}^2\\
C^{(2)} &=& {1\over 8\pi}\sum \ell(\ell+1)(2\ell+1) C_{\ell}
  B_{\ell}^2 F_{\ell}^2,
\end{eqnarray}
where we have used the relation $P_{\ell}'(1) = \ell(\ell+1)/2$
(see equation [5]). The genus per steradian is then
\begin{equation}
g(\nu) = {1\over 2(2\pi)^{3/2}}
  {\sum \ell (\ell+1)(2\ell+1) C_{\ell} B_{\ell}^2 F_{\ell}^2
    \over \sum (2\ell+1)C_{\ell} B_{\ell}^2 F_{\ell}^2} \nu e^{-\nu^2/2}.
\end{equation}
We note again that the genus curve shape is fixed by the Gaussian random-phase
nature of the field, and its amplitude depends only on the shape of the CMB
anisotropy power spectrum and not on its amplitude.

For scale-free energy-density perturbations with matter power spectrum 
$P(k)=\langle|\delta_k|^2\rangle = A k^n$ in a spatially-flat universe, the CMB 
anisotropy angular power spectrum is (Peebles 1982; Bond \& Efstathiou 1987)
\begin{equation}
C_{\ell} = {2^{n-1}A\over (2c/H_o)^{n+3}}
                {\Gamma(3-n)\over\Gamma^2({4-n\over 2})}
                {\Gamma({2\ell+n-1\over 2}) \over
                 \Gamma({2\ell+5-n\over 2})}.
\end{equation}
We can estimate the slope of the CMB anisotropy power spectrum near the 
smoothing scale from the amplitude of the genus curve by using equation (10) 
and this formula.

\subsection{Topology of the MAP Simulations}

When the MAP data becomes available one can measure the genus of the observed
CMB anisotropy and so constrain cosmological models and structure
formation mechanisms. Here we use our simulated CMB anisotropy maps to study
the sensitivity of the genus statistic to models in the presence of instrumental
noise and when the sky coverage is incomplete.

We use the stereographic projection to map the CMB anisotropy sky simulations
onto a plane.  The stereographic projection is conformal and preserves
local geometry.   For a hemisphere the projection is defined by
\begin{equation}
  \rho = 2\tan\left({{\pi/2-|b|}\over 2}\right);~~
  \phi = \ell ,
\end{equation}
where $\rho$ is the radius and $\phi$ is the position angle in the projected
plane, $b$ is galactic latitude and $\ell$ is galactic longitude (and should
not be confused with the Legendre polynomial index). Figure 3 (Plate 2) shows 
such projections for the OCDM and $\Lambda$CDM simulations.
  
When computing the genus we exclude regions with $|b| < 30\deg$ to excise the
Galactic plane region where the Galactic signal is likely to be 
important.\footnote{
When the MAP data becomes available it will be possible to use a less 
restrictive, but more complex, Galactic-plane cut, as was done for the 
DMR data (Banday et al. 1997).}
We use the method of Gott et al. (1990) to compute the topology of the proper
planar projection of the simulated CMB anisotropy sky. In Figure 5 points with
error bars show the two-dimensional genus topology, at $1.0^\circ$ FWHM 
smoothing, derived from the simulated MAP data (accounting for 
noise). The error bars are formal errors in the mean 
derived from the variance among the genus measurements in each
of the eight octants of the sphere.  The curves show the function $g =N\nu
e^{-\nu^2/2}$, the form expected for a random-phase Gaussian field, where $N$
has been adjusted to fit the points optimally via the least-squares method.
For goodness of fit, the points along the curves should be compared with
Student's $t$-variates, rather than Gaussian variates, since the error
bars have been estimated from the data themselves (rather than
independently).  Following Gott et al. (1990) and Colley (1997), we use the 
variable
\begin{equation}
   \label{tilchi}
   \tilde{\chi}^2 = \sum_{i=1}^{21} {{(\bar{g}_i -
   {g_{i, {\rm fit}})^2}\over{{s_i}^2}/(n-1)}}
\end{equation}
to estimate the goodness of fit. The sum runs over the 21 $\nu$ values at which 
the genus $g_{i}$ is computed; $\bar g_i$ is the mean genus among the eight 
octants of the sphere and $s_i{}^2$ is the variance.  We find for the 
OCDM case $\tilde{\chi}^2 = 23.6$, and for FCDM $\tilde{\chi}^2 = 26.3$.  Since
$\tilde{\chi}^2$ does not follow a $\chi^2$ distribution, due to the
non-Gaussian distributions of the $g_i$'s, we generate $10^4$
Monte Carlo runs of $\tilde{\chi}^2$ variables with 20 degrees of freedom, and
8 samplings within each degree of freedom, exactly as we expect in equation
(\ref{tilchi}) (since we have fit one parameter, the amplitude).  In each run,
we generate 8 independent ($\mu = 0, \sigma = 1$) Gaussian variates within 20
independent groups.  From the eight samples in each group $i$ we estimate the
mean ($\bar{g}_i$) and the standard deviation in the mean $s_i/\sqrt{n-1}$; 
$\bar{g}_i/ (s_i/\sqrt{n-1})$ is thus a ($\mu = 0, \sigma = 1$) $t$-variate.  
In the $10^4$ Monte Carlo simulations, we compute
$\tilde{\chi}^2$ as in equation (\ref{tilchi}), by summing the squares of the
twenty $t$-variates.  We find that the $\tilde{\chi}^2$ values of 23.6 and 26.3
fall at the 43\% and 54\% levels of the cumulative distribution of the $10^4$
simulated $\tilde{\chi}^2$ variables, demonstrating an excellent agreement with
the theoretical curve.

In addition to noise, several non-CMB anisotropy foregrounds could contaminate
the MAP data. Perhaps the foreground of most consequence for the genus 
analysis is the radio point-source background (mainly AGNs). To assess its significance, we have added point sources to the $0.3\deg$ FWHM smoothed 
OCDM map. These point sources (also smoothed) are taken to have a number 
distribution in flux, $f$, given by $dN(f) = kf^{-2.5} df$. The normalization 
$k$ is chosen to reflect the real sky and has an average of one source at 
3 $\mu$K or brighter per $0.3\deg \times 0.3\deg$ pixel (Holdaway, Owen, \&
Rupen 1994). Comparing to the OCDM map with noise alone, we find that the 
point sources introduce an rms scatter in $g(\nu)$ of 23 (relative to the 
point-source-free map), significantly smaller than the typical uncertainty in 
$g(\nu)$, which is 42 in the map without point sources. We have also compared 
the genus values (between the maps with and without point sources) using the 
$\chi^2$ statistic. We find $\chi^2= 2.8$ out of an expected 21 (21 $\nu$ 
values), confirming that the effect of the point sources on the genus is very 
small. Since larger smoothing lengths would diminish the signal from point 
sources even further (as with the noise), we conclude that the effects of the 
point-source foreground on the genus are small.

Galactic foregrounds, including warm dust and free-free emission, have been
selected against in choosing the frequency coverage of MAP. At the optimum
frequencies, these foregrounds are expected to be nearly an order of magnitude 
smaller than 
the CMB anisotropy. Furthermore, in the analysis of the DMR data (Bennett 
et al. 1996; G\'orski et al. 1996, and references therein) the $COBE$-DIRBE 
140 $\mu$m sky map (Reach et al. 1995) was used to correct for dust and 
free-free emission contamination, and the 408 MHz all-sky radio survey (Haslam 
et al. 1982) was used to correct for synchrotron emission contamination.
A similar approach, strengthened by new all-sky survey data, should (hopefully)
suffice for dealing with these contaminants in the MAP data.

Another possible foreground is the Sunyaev-Zel'dovich effect in 
clusters of galaxies (Sunyaev \& Zel'dovich 1972). Here hot electrons 
Compton scatter the CMB photons to uniformly higher frequency, introducing an
apparent decrease in brightness temperature as great as $\sim 1$ mK on
the few arc-minute scale (for very massive clusters with a velocity dispersion
of $\sim 1000$ km/s, Markevitch \etal\ 1992).  At smoothing lengths of
$0.3\deg$ and higher, the effect is reduced to $\sim 10\ \mu$K, significantly 
below the peaks in the CMB anisotropy (see Figure 2).  Also, X-ray
surveys of hot gas in the intracluster medium can be used to locate the massive
clusters causing the shift, and to help estimate the expected magnitude of the
shift (e.g., Klein \etal\ 1991). (In the MAP data, the large Sunyaev-Zel'dovich 
signal will come from the Coma cluster; this region can be excised from 
the analysis. This signal can also be estimated by projecting the thermal 
fluctuations out of the two highest frequency MAP channels. A combination of
these cuts should be able to remove most of the Sunyaev-Zel'dovich contamination from the MAP data.)

Non-Gaussian features in the CMB anisotropy will affect the genus curve in many
different ways. Here we consider two cases that can be quantified.
Non-Gaussianity can shift the observed genus curve to the left (towards
negative thresholds) or right, near the mean threshold level. It can also alter
the amplitudes of the genus curve at positive and negative levels differently,
causing $|g(\nu = 1)| \neq |g(\nu = -1)|$. The direction and degree of shift and
asymmetry of the genus curve depend on the number, size and height/depth of hot
and cold spots.  We have computed limits on these non-Gaussianity measures from
our simulated maps.  In the FCDM model we find that the MAP data will detect a
shift of the genus curve near $\nu=0$ (the five $\nu$ values in Figure 5
centered on $\nu = 0$) for $|\Delta\nu| > 0.01\ (0.04)$ for FWHM smoothing of
$0.3\deg\ (1.0\deg)$; the limits are the 2-$\sigma$ error bars
derived from the variances between fits for $\Delta\nu$ from the eight octants.
The asymmetry of the genus amplitude near $\nu=\pm 1$ (the five $\nu$ values
centered on 1 and $-1$) can be detected for $|\Delta g/g(\nu = \pm 1)| > 0.8\%\ 
(4\%)$ at $0.3\deg\ (1\deg)$ FWHM smoothing (also 2-$\sigma$).  
The limits for $|\Delta\nu|$ and $|\Delta g/g(\nu = \pm 1)|$ are very similar 
for OCDM and $\Lambda$CDM.

\subsection{Genus Amplitude as a Check of Cosmological Model}

From Figure 5, it is clear that the amplitude of the genus curves for FCDM and 
OCDM differ significantly. We first consider the genus amplitude for the case 
when instrumental noise is ignored. In the four cosmogonies considered, and in 
the scale-invariant power-law model, we expect from equation (10) a theoretical 
genus per steradian of $g(\nu=1) = 101.0\pm 6.2$ (FCDM), $115.4\pm 6.9$ 
($\Lambda$CDM), $74.3\pm 6.6$ (OCDM), $45.2\pm 3.9$ (RCDM), and $44.4\pm 3.9$ 
($n=1$ scale-invariant power-law model), when the total smoothing scale 
(by the beam and the filter) is $1\deg$
FWHM.  The uncertainty limits here account for cosmic variance, computed from
actual realizations of $a_{\ell}^m$'s. Note that these limits are computed from 
the Galactic-plane-excluded maps. Also, since the genus amplitude does not 
depend on the amplitude of the CMB anisotropy angular power spectrum, these
results are insensitive to the uncertainty in the DMR normalization of the 
models. 

We have also computed the genus directly from simulated maps that include 
receiver noise, for a single realization of each model. (Since all the 
model computations use the same random number seed, the cosmic variance effect
is minimal.) In Figure 6 we have plotted these genus amplitudes
at $\nu = 1$ for the FCDM, OCDM and $\Lambda$CDM models.  For comparison, we
have also plotted the genus for the same cosmogonical simulations but now
ignoring noise.  In the inset, we have zoomed in on the values at $0.3\deg, 0.5
\deg$ and $1.0\deg$ total FWHM smoothing, with 1-$\sigma$ error bars derived 
from comparison of the fits to the genus values within each of
the eight octants (these error bars agree very well with the $\Delta\chi^2 =
1$ level, as expected, \cf\ Press \etal\ 1992).  We have listed in Table 1
the genus values, with statistical uncertainty and expected cosmic variance,
from Figure 6. Note that the $0.3\deg$ column values are subject to finite
pixelization of both the model MAP results and the projected maps (Melott 
et al. 1989). Tests have shown that amplitude ratios between models remain 
very nearly constant when moving to higher resolution in the projected maps.
Hence, it is the finite pixelization of our simulations which 
ultimately limits the resolution for the genus computation.

The three cosmogonies of Figure 6 and Table 1 have significantly different 
genus amplitudes, even when the expected instrumental noise is accounted for.
They can thus be discriminated between on the basis of the genus amplitude.
For instance, at the $0.3\deg$ smoothing scale each model is distinguished
from the other two at greater than 6-$\sigma$ confidence (Table 1 and we
have added all error bars in quadrature). (We again emphasize that the other 
cosmological parameters, such as $h$ and $\Omega_B$, are not allowed to vary in these models.) These results may be directly compared to the genus measurements 
from MAP, as an independent check of the results inferred from the power 
spectrum and correlation function analyses. 

\subsection{Effects of Noise on the Genus}

In order to understand more completely the effects of receiver noise on the
genus, we can look in detail at what is happening, structure by structure, when
noise is added.  At a high temperature cut, for instance, noise can add a hot
spot by adding a small amount of positive temperature to a region near the cut;
conversely, noise can remove a small hot spot by cooling a high-temperature
region to below the cut.  We have illustrated schematically in Figure 7 the
possible effects of noise on the genus. In Figure 7$b$ and 7$e$ the noise has
neither added nor subtracted hot spots, leaving the genus unaltered.  In Figure
7$a$, however, the noise has removed a hot spot and reduced the genus by one;
in 7$c$ the noise has split a hot spot into two and added one to the genus.
In Figure 7$d$ the noise has added an artificial hot spot and one to the
genus while in 7$f$ the noise has merged two hot spots into one and reduced the
genus by one.  These splittings and mergings can occur in higher multiples.

We now examine our simulated maps for these effects.  In Figure 8$a$ we
have plotted the $\nu = 1$ threshold mask for a subsection of the noiseless
OCDM $1.0\deg$ FWHM smoothed map.  In Figure 8$b$ we have plotted the same mask, but for a map with noise added.  Several of the structures in Figure 8$a$ 
become subdivided in Figure 8$b$ --- noise has split hot spots. Also, several
new hot spots appear, while at least one cold spot (in the upper right) develops
within a hot spot.  To understand these changes quantitatively, we have counted
the number of times each case in Figure 7 arises for the OCDM simulations of
Figure 8.  Results from this analysis are given in Table 2.

Table 2 shows that about 80\% (660 out of [78+660+64+15+2+2]) of the structures
counted in the noise-free map correspond to exactly one structure in the map 
with noise. 78 real hot spots occur with no corresponding hot spots in the map 
with noise, while 83 (=64+15+2+2) overlap two or more hot spots in the map with
noise. Very similar correspondence exists in comparing the noise-added map hot 
spots to the noise-free map hot spots, except in the ``0'' column. About a 
quarter (249 out of [249+683+13+2+1]) of the hot spots in the map with noise 
had no corresponding hot spots in the noise-free map.  This is to be expected 
because the noise
adds artificial small-scale bumps on top of the signal.  These bumps can push a
pixel that was just below the threshold (here $\nu = 1$) to a value over the
threshold, so that extra hot spots appear (similarly, extra cold spots appear
at $\nu = -1$).  At $1.0\deg$ FWHM smoothing about two-thirds (683 out of
1011) of the structures in the simulated MAP maps are due to real signal.
Therefore most $1.0\deg$ scale structures in the MAP maps
will correspond to real fluctuations in the microwave background.

\section{Conclusions}

In this paper we have generated mock CMB anisotropy maps of representative
DMR-normalized CDM cosmogonies appropriate for the MAP satellite experiment. 
We have studied the sensitivity of the simulated MAP data to cosmology,
sky coverage, and instrumental noise. An accurate knowledge of instrumental 
noise is essential if the statistical tests considered here are to live up
to expectations. 

We have focused on the correlation function and the genus statistic
as tests of cosmogonical models. If the underlying theory has
Gaussian random phases, then these two statistics are just integrals
of the power spectrum.  However, since they are computed from the
data in a very different manner than the power spectrum, they will
be useful checks on power spectrum computations even in Gaussian random
phase models.  The genus statistic will also be a powerful test of the
random phase hypothesis. 

The correlation function statistic provides an accurate measurement of the 
acoustic Hubble radius at decoupling from the simulated MAP data. The 
$\Omega_0 = 0.4$
OCDM model considered here can be distinguished from the spatially-flat FCDM
and $\Lambda$CDM models with more than 99\% confidence from the location of the
acoustic valley in the correlation function, even when the expected
amount of instrumental noise is present.  The FCDM and $\Lambda$CDM models
can also be distinguished at separations $\sim 0.5\deg$ by the amplitude
of the correlation function.  Note however that for these models all other
cosmological parameters (such as $\Omega_B$ and $h$) are fixed.

The genus of the CMB fluctuations can also be accurately measured from the MAP
data.  The genus curves of Gaussian cosmological models allow a 2-$\sigma$ statistical
horizontal shift of the zero-crossing point by only $|\Delta\nu| \la 0.01\ 
(0.04)$ when the total effective smoothing is $0.3\deg\ (1\deg)$ FWHM.
The asymmetry of the genus curve at the positive and negative 
threshold levels is only $|\Delta g/g(\nu = \pm 1)| \la 0.8\%\ (4\%)$ in the 
Gaussian cosmogonies considered. Deviations of the observed 
MAP data that exceed these values will be evidence for non-Gaussian behavior.

The amplitude of the genus curve, which is a measure of the shape of the power
spectrum at the smoothing scale, can also be a powerful discriminator between
cosmological models.  Even with the expected amount of instrumental noise and
partial sky coverage, the MAP data should discriminate between the
cosmogonies considered here at confidence levels exceeding 99\% just
on the basis of the genus amplitude analysis.

Of possible outcomes from the MAP experiment, a more interesting one
would be if the results were in close agreement with the low-$\Omega_0$
open-bubble inflation model. Not only would this tell us the value of
$\Omega_0$, but it would affirm inflation and support the open-bubble 
inflation model in particular, thus providing evidence that there are universes 
other than our own (other bubble universes), all arising out of an initial 
metastable inflation state.

\acknowledgments

CBP acknowledges the support of the Basic Science Research Institute Program 
BSRI97-5408, and the KOSEF program 96-1400-04-01-3. JRG and WNC are supported 
by NSF grant AST-9529120 and by NASA Grant NAG5-2759.  WNC is grateful to the 
Fannie and John Hertz Foundation for its gracious support. This paper represents part of WNC's thesis at Princeton University. BR acknowledges 
support from NSF grant EPS-9550487 with matching support from the state of 
Kansas and from a K$^*$STAR First award. DNS acknowledges the MAP/MIDEX project
for support.

\clearpage

\begin{table}
\begin{center}
\caption{Numerical values for genus amplitude per steradian at 
$\nu = 1$\tablenotemark{a}}
\vspace{0.3truecm}
\tablenotetext{{\rm a}}{Computed at various smoothing scales for the
FCDM, OCDM and $\Lambda$CDM models, from simulated maps that account for the 
expected instrumental noise and the Galactic-plane cut. The first uncertainty 
listed is the statistical error of the fit (1-$\sigma$, derived from comparing 
the genus in eight octants of the sphere), and the second uncertainty is that 
due to cosmic variance.}
\begin{tabular}{lccc} 
\tableline\tableline
Smoothing scale & FCDM & OCDM & $\Lambda$CDM \\
\tableline
$0.3\deg$ & $3650 \pm 16 \pm 23$
          & $3995 \pm 21 \pm 40$
          & $3313 \pm 14 \pm 22$ \\
$0.5\deg$ & $637.6 \pm 6.0 \pm 13.6$
          & $719.1 \pm 8.2 \pm 22.1$
          & $617.9 \pm 5.2 \pm 14.0$\\
$1.0\deg$ & $137.7 \pm 2.5 \pm 6.2$
          & $120.9 \pm 2.6 \pm 6.6$
          & $149.7 \pm 2.5 \pm 6.9$ \\
\tableline
\end{tabular}
\end{center}
\end{table}
%

\begin{table}
\begin{center}
\caption{Effects of noise on the genus\tablenotemark{a}}
\vspace{0.3truecm}
\tablenotetext{{\rm a}}{Also see Figures 7 and 8. At the threshold
$\nu = 1$, for a given hot spot in either the noise-added map or noise-free
map, we have counted the number of hot spots in the other map, which that hot
spot intersects.  In the first row, we have the number of times a hot spot in
the noise-free map intersects a hot spot in the map with noise: noise can
cause a hot spot to disappear (col. 0; Figure 7$a$); noise might leave a hot
spot as is (col. 1; Figure 7$b$); noise might split a hot spot in twain
(col. 2; Figure 7$c$), etc.  In the second row, we have the number of times a
hot spot in the map with noise intersects a hot spot in the noise-free map:
the noise might have produced a whole new hot spot (col. 0; Figure 7$d$); noise
could have left one spot as one spot (col. 1; Figure 7$e$); noise could have
merged two hot spots into one (col. 2; Figure 7$f$), etc.  For example, 78 hot
spots in the noise-free map are eliminated in the map with noise,
while 660 of them are left as is, and 64 are split into two spots.  Similarly,
249 of the hot spots in the map with noise are eliminated in the noise-free map,
while 683 of them are left as is, and 63 are split into two.  The change of
genus associated with these effects is given in the last row.}
\begin{tabular}{lccccccc} 
\tableline\tableline
& \multicolumn{7}{c}{intersecting hot spots} \\
map$_1$ $\rightarrow$ map$_2$ & 0 & 1 & 2 & 3 & 4 & 5 & 6 \\
\tableline
noise-free $\rightarrow$ noise &
 78 & 660 &  64 &  15 &   2 &   2 &   0 \\
noise $\rightarrow$ noise-free &
249 & 683 &  63 &  13 &   2 &   1 &   0 \\
\tableline
change of genus & -1 & 0 & 1 & 2 & 3 & 4 & 5 \\
\tableline
\end{tabular}
\end{center}
\end{table}

\clearpage

\begin{figure}
\caption{Curves show the CMB anisotropy angular power spectra, as a 
function of multipole $\ell$, of the four models considered: FCDM (solid),
${\Lambda}$CDM (dotted), OCDM (dashed), and RCDM (long dashed). $T_0$ is the mean CMB temperature. The clouds of dots, tracing the curves at small $\ell$, 
are from actual realizations of the $a_{\ell}^m$'s with $0.21\deg$ FWHM beam 
smoothing as appropriate for MAP, and show the cosmic variance at each $\ell$.
The clouds of dots in panel ($a$) do not account for instrumental noise, 
while those in panels ($b$) and ($c$) do. The dots in the last two panels 
correspond to $0.3\deg$ and $1.0\deg$ total smoothing of the signal, 
respectively.}
\end{figure}

\begin{figure}
\caption{Upper panel is a Mollweide projection of the CMB anisotropy
map simulated for the OCDM model. The map has been convolved by a Gaussian 
filter with $0.3\deg$ FWHM. Lower panels are magnifications of the 
Mollweide projection of the simulated CMB anisotropy maps for the OCDM and 
$\Lambda$CDM models. In each lower panel map the equator runs horizontally through the center, where the plots cover $30\deg$ horizontally and $24.4\deg$
vertically.}
\end{figure}

\begin{figure}
\caption{Upper panel is the OCDM CMB anisotropy map, in 
stereographic projection and with $1\deg$ FWHM Gaussian smoothing, 
showing the north and south galactic caps ($|b| > 30\deg$). The projection is 
conformal and the shape of structures are locally undistorted. Lower panel 
is a similar stereographic projection for the $\Lambda$CDM model. For the 
purpose of this comparison both models have identical phases, in both 
signal and noise (picked from the same random number seeds).}
\end{figure}

\begin{figure}
\plotfiddle{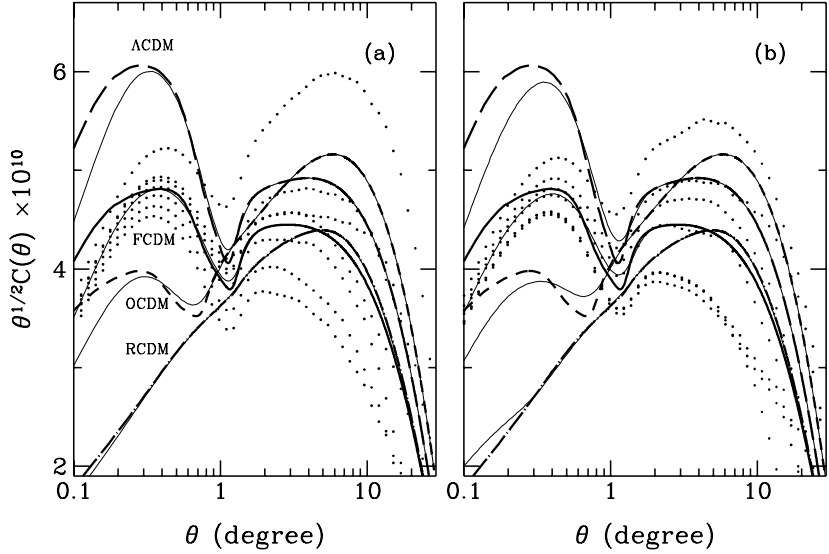}{12cm}{0}{60}{60}{-185}{-90}
\caption{Correlation functions of the four cosmogonies. Thick curves are the 
correlation functions transformed from the power spectra shown with curves in 
Figure 1. Thin solid curves, tracing the thick ones at large angular 
separations, account for the convolution with a $0.21^\circ$ Gaussian beam in 
panel ($a$), with an additional $0.214^\circ$ Gaussian smoothing in panel 
($b$). Dotted curves  are the correlation functions
from seven independent realizations of FCDM CMB anisotropy 
skies showing the cosmic variance. Thin solid and dotted curves in panel 
($a$) ignore instrumental noise, while those in panel ($b$) account for it. 
The Galactic-plane cut is not accounted for. Clearly, inclusion of 
instrumental noise smoothed over a small scale ($\sim$ beam size) does not 
change the correlation function much, indicating that the statistical
uncertainty is dominated by cosmic variance.}
\end{figure}

\begin{figure}
\plotfiddle{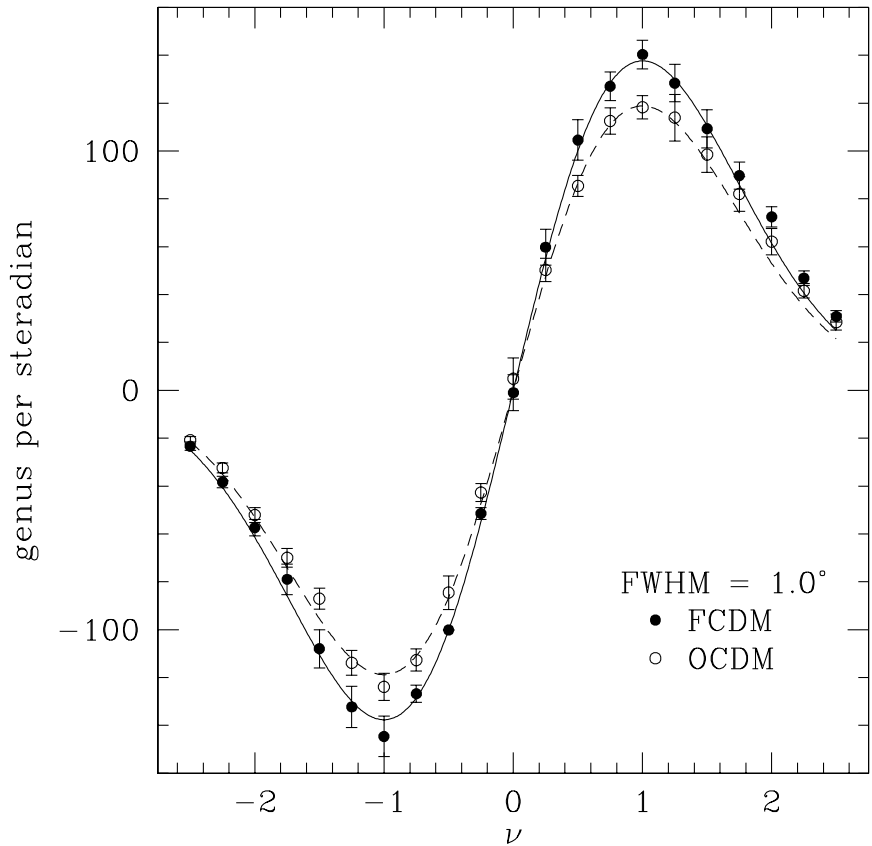}{12cm}{0}{60}{60}{-185}{-90}
\caption{
Two-dimensional genus topology from the MAP simulations accounting for 
instrumental noise, with $1.0^\circ$ FWHM smoothing and
for $|b| > 30\deg$.  Open symbols are for OCDM, closed for FCDM. Error bars
are derived from the variance of the genus computed independently in each
of the eight octants of the sphere.  The curves show the functional form
expected for a random-phase Gaussian field, $N\nu e^{-\nu^2/2}$, and have been
fit to the points by adjusting $N$.  The cosmogonies can clearly be
distinguished from each other.}
\end{figure}

\begin{figure}
\plotfiddle{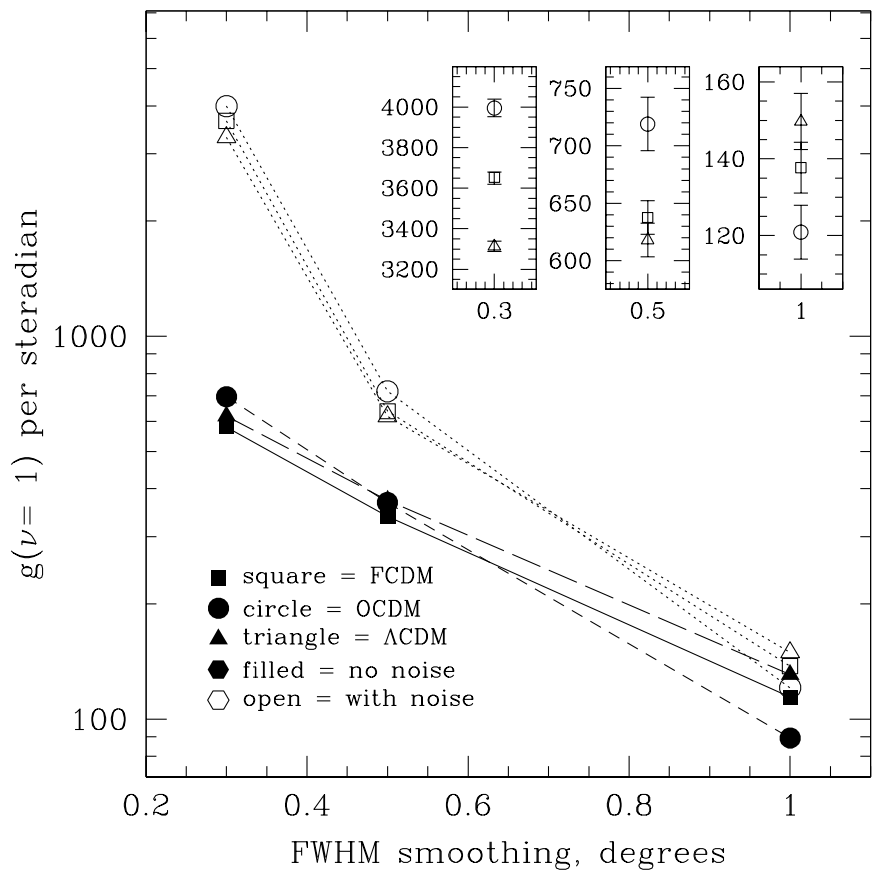}{12cm}{0}{60}{60}{-185}{-60}
\caption{Fit amplitudes (at $\nu = 1$) of genus curves for various cosmogonies, 
as in Figure 5.  Filled squares are results from noise-free maps for
FCDM, filled triangles for $\Lambda$CDM, and filled circles for OCDM.
Open squares are results from maps with noise for FCDM, open triangles
for $\Lambda$CDM, and open circles for OCDM.  The error bars shown in the 
insets are the quadrature sum of the statistical uncertainty and that 
due to cosmic variance (see Table 1).}
\end{figure}

\begin{figure}
\plotfiddle{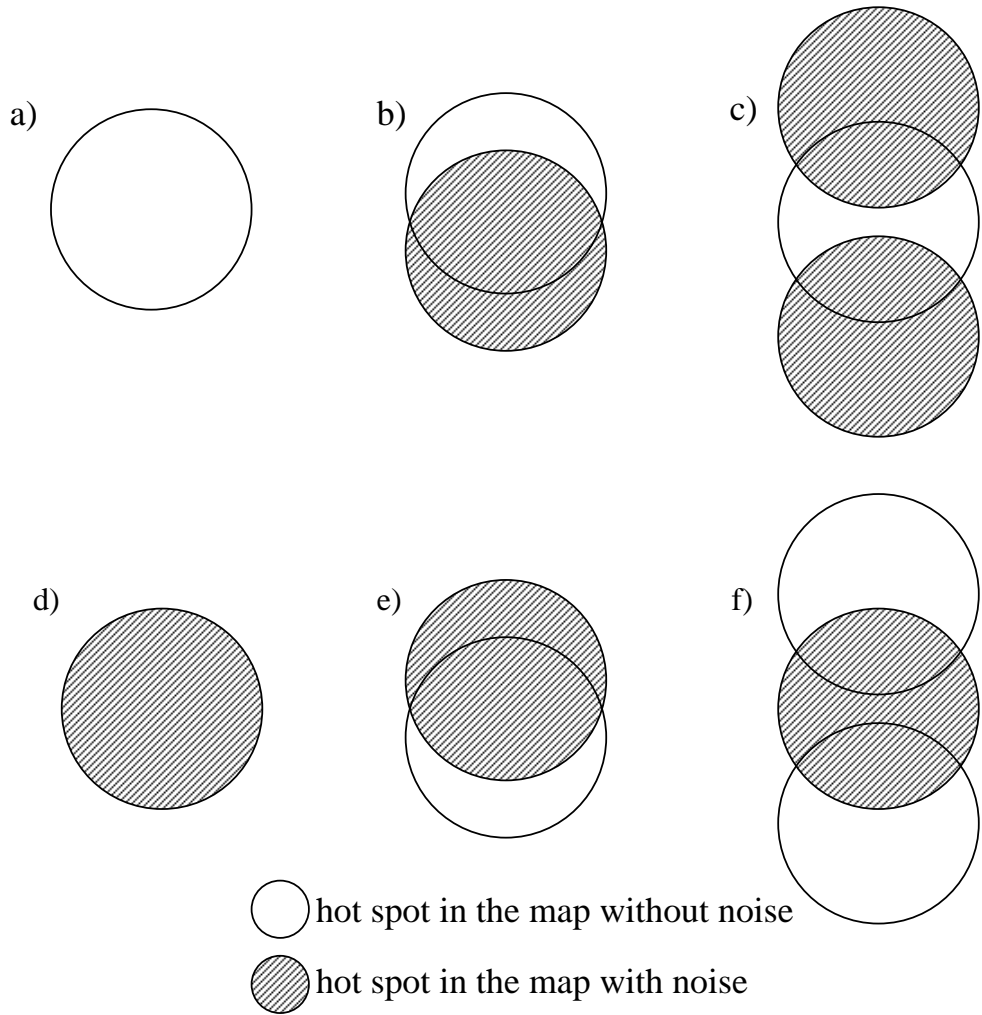}{12cm}{0}{120}{120}{-210}{-500}
\caption{Overlays of simulated maps with and without instrumental noise.
This schematic illustrates the effect of noise on the genus by comparing 
how well a given structure in the map with noise corresponds to a structure in 
the noise-free map.  If a structure in the map with noise, such as a hot spot, 
intersects exactly one hot spot in the noise-free map, as in cases $b$) and 
$e$), then the noise has not altered the genus.  
However, if the noise is significant, artificial hot spots can appear or real 
hot spots can disappear, both of which would alter the genus.  $a$) is a case 
where the noise has removed a hot spot (there is a hot spot in the noise-free 
map but not in the map with noise), $c$) where the noise has split a hot spot 
into two.  $d$) is a case where the noise has added a hot spot (there is a hot
spot in the map with noise but not in the noise-free map), 
$f$) where the noise has merged two hot spots into one.}
\end{figure}

\begin{figure}
\plotfiddle{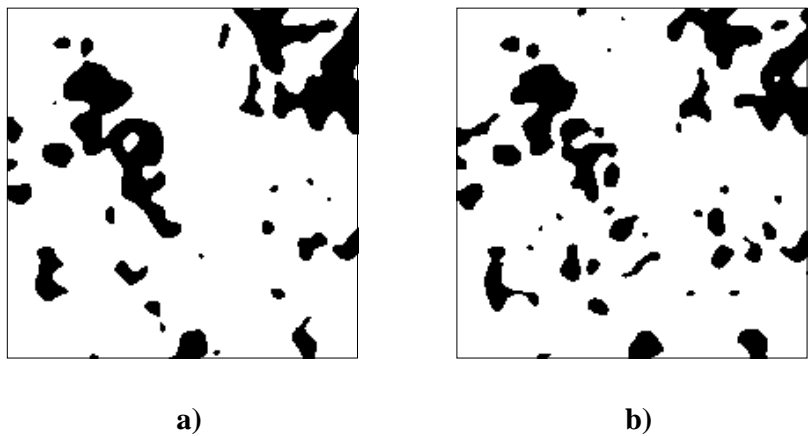}{12cm}{0}{60}{60}{-185}{-90}
\caption{Threshold masks from a subset (approximately $30\deg \times 
30\deg$) of the OCDM CMB anisotropy simulations at $1.0\deg$ FWHM 
smoothing: ($a$) ignoring instrumental noise, ($b$) accounting for instrumental
noise (rms noise $= 35\ \mu$K per $0.3\deg \times 0.3\deg$ area).}
\end{figure}

\end{document}